\documentclass{ws-rv9x6}
\usepackage{ws-rv-van}             % numbered citation/references (default)
\makeindex
\newcommand{\be}{\begin{equation}}
\newcommand{\ee}{\end{equation}}
\newcommand{\bea}{\begin{eqnarray}}
\newcommand{\eea}{\end{eqnarray}}
\newcommand{\lQ}{\Lambda_{\rm QCD}}
\newcommand{\als}{\alpha_{\rm s}}
\begin{document}

\setcounter{chapter}{1}
\chapter*{Non-relativistic bound states: 
the long way back from the Bethe--Salpeter to the Schr\"odinger equation$^1$}{\footnotetext[1]{Contribution
to ``Fundamental Interactions---A Memorial Volume for Wolfgang Kummer'',
D.\ Grumiller, A.\ Rebhan, D.V.\ Vassilevich (eds.)}}
%\chapter[Non-relativistic bound states]{Non-relativistic bound states: 
%the long way back from the Bethe--Salpeter to the Schr\"odinger equation}
\author[Antonio Vairo]{Antonio Vairo}
\address{Physik Department, Technische Universit\"at M\"unchen, 85748 Garching, Germany \\ 
E-mail: {\tt antonio.vairo@ph.tum.de}}

\begin{abstract}
I review, in a personal perspective, the history of the theory of
non-relativistic bound states in QED and QCD from the Bethe--Salpeter equation 
to the construction of effective field theories. 
\end{abstract}

\body

\section{Introduction}
The study of bound states and, in particular, of non-relativistic bound states 
has accompanied the quantum theory from its beginning through all its subsequent turning points 
up to what is now the Standard Model of particle physics. 
At the beginning it was the description of the hydrogen atom 
that led to the foundation of quantum mechanics, later  
the Lamb shift contributed to the development of 
relativistic field theories and renormalization, which 
eventually led to the foundation of Quantum Electrodynamics 
(QED); similarly, in the seventies, quarkonium  played a special role in 
the foundation of Quantum Chromodynamics (QCD).
The special role of non-relativistic bound states in particle physics 
is due to the striking experimental signatures that they provide 
and the fact that analytical (perturbative) methods 
are able to describe the relevant features of these signatures.

Despite this, it has proven very difficult to carry out theoretical analyses 
of a precision comparable with the data, in part due to the high quality 
of the data, but largely owing to the difficulties in performing 
bound-state calculations. These may be traced back to the presence  
of different energy scales that make it a challenge to maintain a 
consistent book-keeping in the calculations. 

\index{Schr\"odinger equation}
Let us consider a non-relativistic particle of mass $m$ that propagates in a potential $V$
(in the case of a Coulomb potential: $V = -\alpha/r$). 
If the momentum of the particle is non relativistic, then $p\sim m v$, $v \ll 1$ being the velocity
of the particle. In the threshold region, the velocity is such that $m v^2 \sim V$.
The balance between kinetic energy and potential creates the bound state: 
the  particle propagator $G$ cannot be computed order by order in $V$, 
but comes from resumming all potential insertions in the free propagator $G_0 = 1/(E- p^2/2m)$:
\be
G = G_0 + G_0VG .
\label{schroeG}
\ee
The function $\displaystyle G = 1/(E-p^2/2m-V)$ 
exhibits poles in correspondence of the bound-state energies $E_n$ 
$\sim m v^2$ ($= - m \alpha^2/2n^2$ in the Coulombic case, which implies $v \sim \alpha$ and 
$1/r \sim m \alpha$), the residues at the poles, $\phi^*_n\phi_n$, satisfy the equation:
\be
E_n \, \phi_n = \left( \frac{p^2}{2m}+  V\right)\,\phi_n, 
\label{schroephi}
\ee
which is the Schr\"odinger equation for a non-relativistic bound state 
whose wave function is $\phi_n$.

Hence, the non-relativistic dynamics of a particle close to threshold is characterized  by a hierarchy of 
energy scales: $m \gg mv \gg mv^2$. The scale of the mass is sometimes called ``hard'', 
the scale of the typical momentum transfer, or inverse size of the system, $mv$, is called ``soft''
and the scale $mv^2$ is called ``ultrasoft''.

At the level of non-relativistic quantum mechanics, 
$m$ does not play any dynamical role, for the kinetic energy is $p^2/2m$ rather than 
$\sqrt{p^2+m^2}$. The contributions from the other scales are accounted for by the Schr\"odinger equations 
(\ref{schroeG}) or (\ref{schroephi}). The solutions of the Schr\"odinger equation are non-relativistic bound states  
of typical energy of order $mv^2$ and typical momentum (or inverse size) of order $m v$.

One may expect that a more complicated picture will emerge in a relativistic 
field theory, although the leading dynamics should be still described by a Schr\"odinger equation.
In a relativistic field theory description of the bound state, we will have, besides the bound state, 
other degrees of freedom, for instance photons (in QED) and gluons (in QCD) emitted 
and exchanged by the bound state; for each of them, modes associated to each of 
the energy scales, $m$, $mv$ and $mv^2$ will appear. We shall discuss bound states in 
relativistic field theories in the next section.

\section{The Bethe--Salpeter equation}
\index{quarkonium}
\index{positronium}
Let us consider a particle and an antiparticle (e.g. an electron and a positron or a quark and 
an antiquark) that interact near threshold. In the centre-of-mass frame, their momenta $p$ and energies $E$ 
are small compared to their masses $m$: $p/m \sim v \ll 1$. 
We assume that we may express the interaction perturbatively in terms of Feynman diagrams.
This is always the case in QED, but does not need to be so in QCD where, at the typical 
hadronic scale $\lQ$, perturbation theory breaks down. Non-relativistic bound states 
in QCD are made by heavy quarks: this means that at least the 
quark mass is larger than $\lQ$. A bound state of a heavy quark and a heavy antiquark 
is called quarkonium (examples are charmonium, a charm-anticharm bound state, 
and bottomonium, a bottom-antibottom bound state; a top-antitop bound state, which would be 
toponium, has no time to form due to the rapid top quark weak decay, however, near threshold, the bound-state   
enhancement should be visible in the top-antitop production cross section). A perturbative treatment 
of quarkonium, which requires $mv$, $mv^2$ $\gg \lQ$, is justified only for
top-antitop pairs near threshold and possibly for the ground state of bottomonium. 

\begin{figure}[htb]
\makebox[1truecm]{\phantom b}
\put(20,0){\epsfxsize=5truecm \epsfbox{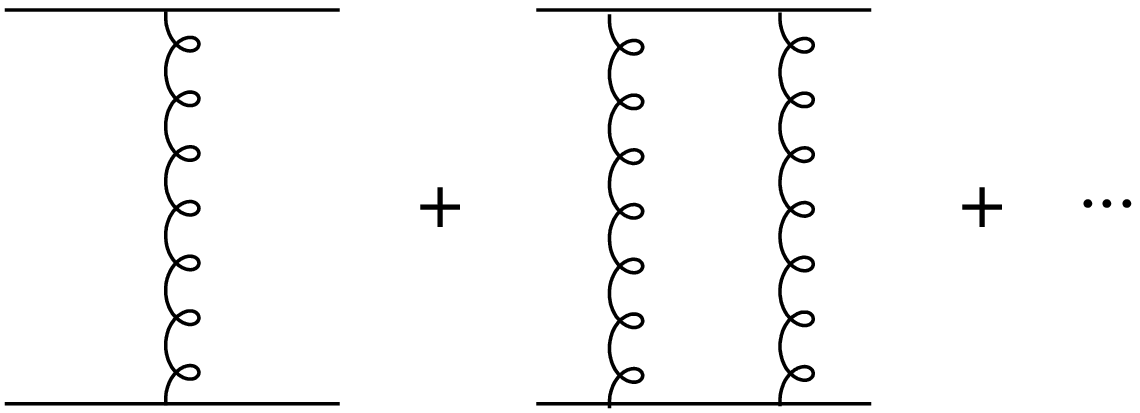}}
\put(180,25){$\displaystyle \approx \frac{1}{E - {p^2}/{m} - V}$}
\put(38,-19){$\als\left(1+\als/v + \dots \right)$ }
\caption{Resummed propagator near threshold.
\label{figcoulomb}}
\end{figure}

How does the bound state emerge in a near threshold interaction?
For certain sets of graphs, like those in Fig. \ref{figcoulomb}, 
the perturbative expansion breaks down when $\als \sim v$ (for definiteness, 
we will consider here and in the following figures the QCD case: continuous lines 
stand for quarks and antiquarks, and the curly lines for gluons; the strong coupling 
constant is $\als$). 
The summation of all $\als/v$ contributions leads
to the appearance of a bound-state pole of order $mv^2\sim m\als^2$ 
in the resummed propagator.  Indeed, in the leading non-relativistic limit, 
when the quark/antiquark propagators can be approximated by
$\displaystyle \frac{i}{\pm p^0+E/2 - {\bf p}^2/2m + i\epsilon}\frac{1\pm \gamma^0}{2}$ 
and the gluon exchange by $\displaystyle \frac{i}{{\bf q}^2}$
(close to threshold we may expand in $|q^0|/|{\bf q}| \ll 1$; $\gamma^0$ is a
Dirac matrix) the Green's function shown in Fig. \ref{figcoulomb} satisfies Eq. (\ref{schroeG}). 

\begin{figure}[htb]
\makebox[2truecm]{\phantom b}
\put(20,0){\epsfxsize=6truecm \epsfbox{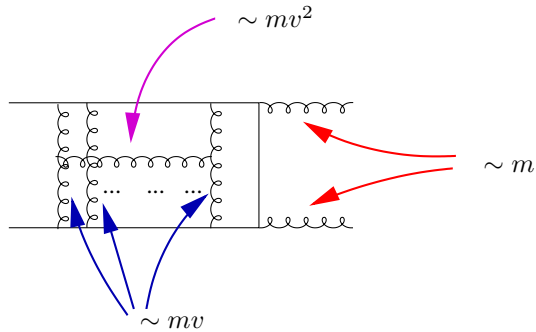}}
\put(70,-4){$\sim mv$}
\put(107,110){$\sim mv^2$}
\put(200,55){$\sim m$}
\caption{Annihilation diagram contributing to the quarkonium decay width.
\label{figentangled}}
\end{figure}

Beyond the leading non-relativistic limit, diagrams will be  
much more complicated to calculate and contributions from the 
different energy scales will get entangled. This happens for any diagram, but 
the annihilation diagram shown in Fig. \ref{figentangled} provides 
a rather immediate way to see it. 
Assuming that the incoming quarks are near threshold, the different gluons entering 
the diagram are characterized by different scales: the annihilation gluons 
have a typical energy of order $m$; 
binding gluons carry the momentum of the incoming quarks, 
which is of order $mv$, and ultrasoft gluons, sensitive to the intermediate 
bound state, have energies of the order of the binding energy, i.e. $mv^2$.

\index{Bethe--Salpeter equation} 
The entanglement of the different energy modes makes it difficult to organize 
a full relativistic calculation of the bound state in QED or QCD. An equation
suitable for bound states in field theory, formally
similar to (\ref{schroeG}), was suggested almost sixty years ago by H. Bethe and
E. Salpeter \cite{Bethe:1951aa}:
\be 
G = G_0 + G_0KG , 
\label{BS} 
\ee
where $G$ is the two-particle Green's function, $G_0$ the product of the 
free propagators of the two particles and the kernel $K$ is the sum of all 
amputated irreducible two-particle diagrams. 
Equation (\ref{BS}) does not represent an expansion, because, like (\ref{schroeG}), a
bound state emerges only from the sum of all interactions, at least those shown  in Fig. \ref{figcoulomb}.
However, unlike (\ref{schroeG}), 
the Bethe--Salpeter equation is not homogeneous in the momentum scale and an exact solution is unknown. 
To make the Bethe--Salpeter equation useful, the strategy has been to 
isolate from $K$ a kernel $K_c$ containing the leading contribution responsible 
for the formation of the bound state, i.e. the Coulomb potential, and expand around 
it (see, for instance, Ref. \refcite{Lepage:1977gd}). The most refined approach in this strategy 
can be found in Ref. \refcite{Barbieri:1978mf} (see also Ref. \refcite{Vairo:1998gy}): 
$K_c$ is chosen in such a way that the corresponding Bethe--Salpeter equation, 
$G_c = G_0 + G_0 K_c G_c$, may be solved in an analytically closed form, 
and the full Green's function expanded around the exact solution:
\be
G = G_c + G_c \delta K G,
\label{BSc}
\ee
where $\delta K  = K -K_c$. Since, in the non-relativistic limit, $K_c$
becomes the Coulomb potential, $G_c$ is nothing else than a relativistic
modification of the solution of Eq. (\ref{schroeG}) for $V$ equal to the Coulomb potential, 
which is known since long time \cite{Schwinger:1964zzb}.
The difference between Eq. (\ref{BS}) and Eq. (\ref{BSc}) is that the latter is a perturbative 
expansion in the kernel while the former is not.

The Bethe--Salpeter equation was the only systematic tool to treat bound states in field theory 
until the end of the eighties. However, around that time, it became increasingly clear that 
perturbative calculations for QED bound states, to which the Bethe--Salpeter equation had been mostly applied, 
could not be push beyond the reached limit if not at the cost of a formidable amount 
of work. It shows the difficulty of the approach the fact that going from the calculation of the 
$m\alpha^5$ correction in the hyperfine splitting of the positronium ground state \cite{Karplus:1952wp} to the 
$m\alpha^6\ln \alpha$ term \cite{Caswell:1978vz,Bodwin:1977ut} took twenty-five years!  
The main problem was the lack of an efficient 
way of disentangling the contributions coming from the different energy scales and organize 
them in a perturbative expansion (techniques for asymptotic
expansions of Feynman integrals near threshold would be developed later \cite{Beneke:1997zp}): 
each Feynman diagram would contribute to the observables with a series in the coupling constant. 
No obvious counting rules were available even for the 
leading term of the series. Also gauge invariance did not provide a useful organizational tool, 
since it was very cumbersome to isolate gauge-invariant subsets 
of diagrams in $K$ \cite{Love:1977hj,Feldman:1979kw,Feldman:1979kx,Vairo:1995sk}.

In the late seventies and eighties, systematic calculations of quarkonium observables started (for a recent review 
see Ref. \refcite{Brambilla:2004wf}). The complicated dynamics of QCD made it more apparent 
that a treatment based on the Bethe--Salpeter equation was inadequate to perform high-precision 
quarkonium calculations. First, not all of the quarkonium scales are in general perturbative, lower ones 
may not be, so that a separation of scales is necessary to achieve factorization. Second, 
even if a perturbative treatment would be possible (like for the bottomonium ground state 
and for $t\bar{t}$ threshold production), the number and topology of diagrams makes 
the calculation prohibitive. It was felt that somehow going back to the Schr\"odinger equation 
and identifying a quarkonium potential would lead to a more treatable problem. In Refs. 
\refcite{Gupta:1981pd,Gupta:1982qc,Buchmuller:1981aj,Pantaleone:1985uf,Titard:1993nn},
a quarkonium potential was derived from the quark-antiquark scattering amplitude. In the same years, 
focusing in particular on toponium and $t\bar{t}$ threshold physics, 
a similar program was carried out by W. Kummer and collaborators  
\cite{Kummer:1981qc,Kummer:1994bq,Modritsch:1994hv,Kummer:1995id,Kummer:1996jz} 
(see also the Ph.D. thesis  in Ref. \refcite{Modritsch:1995jw}). 
In this case, the starting point was the Bethe--Salpeter equation and the generalization to QCD of 
the solution of the Bethe--Salpeter equation for positronium found in Ref.  \refcite{Barbieri:1978mf}.
Still, the goal was not the solution of the Bethe--Salpeter equation itself, 
but the derivation of a potential, facing, in the process, some of the
problems that, in a few years, would have led to (and found a solution with)  
the construction of effective field theories for non-relativistic bound states.
Among the problems mentioned or addressed at that time were 
the infrared sensitivity of the potential, the inclusion of a finite decay
width (in Refs. \refcite{Modritsch:1994hv,Kummer:1995id}, one can find
addressed, for the first time in a formal way, how to include the top-quark instability
beyond leading order), gauge invariance. 
The infrared sensitivity of the potential will be discussed in Sec. \ref{secpNR}.

\section{NRQED/NRQCD}
\index{NRQED}
\index{NRQCD} 
In QED and QCD, one may take advantage of the hierarchy of scales that
characterizes non-relativistic bound states by expanding Green's functions 
in the ratios of low energy scales over large energy scales. Working out these 
expansions, however, turns out to be cumbersome and does not lead 
to a straightforward and easy way to organize the calculation. If such an expansion is instead  
implemented at the Lagrangian level, it leads to the construction of an effective field theory (EFT).
In the effective field theory the large scale is integrated out from the beginning and does not 
appear anymore in the Green's function. The terms in the EFT Lagrangian are organized as an expansion 
in powers of the inverse of the large scale that has been integrated out leading to a straightforward 
power counting. 

The first EFT introduced for non-relativistic bound states in QED and QCD has been non-relativistic QED/QCD 
(NRQED/NRQCD)\cite{Caswell:1986ui}.
The large scale that is integrated out in NRQED/NRQCD is the mass $m$ of the bound-state constituents.
The degrees of freedom of  NRQED/NRQCD are non-relativistic fermions and antifermions, and 
photons/gluons of energy and momentum smaller than $m$; they build up the
operators $O_n$ of the Lagrangian. The Lagrangian is organized as an expansion in $1/m$:
\be
{\cal L}_{\rm NRQED/NRQCD} = \sum_n c_n(\als(m),\mu) \times \frac{O_n(\mu)}{m^n}.
\label{NRQCD}
\ee
Since, once $O_{n}$ has been run down to energies lower than $m$, the expectation 
value of $O_{n}$ scales like $mv$ or smaller scales, Eq. (\ref{NRQCD}) provides, for any physical observable,  
a perturbative expansion in the ratio of the scale $mv$ or smaller scales over $m$.
The Wilson coefficients $c_n$ are non analytical in the scale $m$ and function of the factorization scale $\mu$.
They are calculated by equating, ``matching'', amplitudes in QED/QCD with
amplitudes in NRQED/NRQCD order by order in $1/m$ and in the coupling constant 
since in both theories we have that $\alpha$, $\als(m) \ll 1$.
The matching may be performed on scattering amplitudes, hence in a manner
completely independent of the bound state. 
This is not surprising: the formation of the bound state takes place at a scale, $mv$, 
which is much smaller than $m$.

\begin{figure}[htb]
\makebox[-0.1cm]{\phantom b}
\put(0,0){{\epsfxsize=10truecm\epsffile{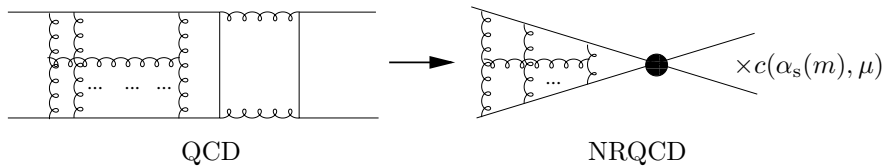}}}
\put(66,-15){QCD}
\put(220,-15){NRQCD}
\put(275,18){$\times c(\als(m),\mu)$}
\caption{Matching to NRQCD.
\label{figNRQCD}}
\end{figure}

The diagram in Fig. \ref{figentangled} corresponds, via the optical theorem, to the imaginary 
part of the diagram shown on the left in Fig. \ref{figNRQCD}. The same process would be described in 
NRQCD by the diagram shown on the right in Fig. \ref{figNRQCD}, i.e. by a diagram where the two 
hard gluons coming from the annihilation are replaced by a contact interaction. 
The difference between the two diagrams is compensated by the Wilson coefficient $c \sim \als(m)^2$.

\index{quarkonium}
As our example may suggest, NRQCD is particularly well suited to describe 
heavy quarkonium decay and production\cite{Bodwin:1992ye,Bodwin:1994jh}.
It is in the theory of quarkonium production that NRQCD has perhaps achieved
its major success by explaining, in the nineties, the quarkonium production data at the Tevatron by a new mechanism
allowed by the symmetries of NRQCD, the octet mechanism, but missed by previous approaches 
(see Ref. \refcite{Brambilla:2004wf} and references therein).

Applications of NRQED have started in the nineties and with time have led to 
many new results (for some early works, see Refs. \refcite{Labelle:1993tq,Labelle:1994ve,Kinoshita:1995mt}).
However, the progress in high precision calculations in NRQED/NRQCD has been slowed down 
by two major shortcomings: first, the fact that soft and ultrasoft degrees of freedom 
still remain entangled in NRQED/NRQCD, second, the use in early NRQED/NRQCD calculations 
of a cut-off regularization scheme. The first difficulty led to a 
power counting that was non homogeneous and to perturbative calculations that still involved 
two scales. To overcome this difficulty, lower energy EFTs were developed; 
we shall discuss some of them in Sec. \ref{secpNR}. 
The second difficulty, on one hand, pushed the development of 
lattice NRQCD \cite{Thacker:1991bm} (see Ref. \refcite{Gray:2005ur} for recent results
on the bottomonium spectrum), on the other hand, addressed analytical studies 
towards a consistent formulation of NRQCD in dimensional regularization.

\section{The bound state in dimensional regularization}
\index{HQET} 
\index{dimensional regularization} 
Surprisingly, it was only few years after NRQCD had been introduced that an EFT for mesons made of 
a single heavy quark, the heavy quark effective theory (HQET), was formulated \cite{Isgur:1989vq}.
In the two-fermion  sector, the Lagrangian of HQET contains the same operators as the NRQCD Lagrangian. 
However, HQET is a quite different theory from NRQCD: HQET contains 
only a single dynamical scale, $\lQ$, which governs its power counting. As a consequence, 
the kinetic energy, which is of order $\lQ^2/m$, is suppressed with respect to the binding energy, 
which is of order $\lQ$, while, in a non-relativistic bound state, the two are of the same order.

It is precisely because, in HQET, propagators are expanded in the kinetic energy that we may 
use dimensional regularization in loop calculations. This has led to a rapid, vast   
and very successful use of the HQET in precision studies of $D$ and $B$ mesons\cite{Neubert:1993mb}.
Instead, keeping the kinetic energy in the denominators of the propagators, as the power counting of NRQCD 
seems to suggest, turns out to be disastrous and leads to the break down of the power counting. 
The reason is that, in dimensional re\-gu\-larization, integrals are not cut-off at high momenta and 
hard scale poles are going to contribute if present in the denominators.
Once, this had been realized in Ref. \refcite{Manohar:1997qy}, it became also
clear that the way out was to compute the matching to NRQCD in the same way as
the matching to the HQET, i.e. order by order in $1/m$. 
Since both in NRQCD and in the HQET the matching conditions are computed in the same way, the two 
Lagrangians are the same: not only the operators of the two theories coincide in the 
two-fermion sector, but also their matching coefficients do. 
Obviously, in order to compute observables with the NRQCD Lagrangian, the usual non-relativistic 
power counting rules, different from the HQET ones, should be used.

Having understood how to treat the bound state in dimensional regularization, 
opened, finally, the doors to analytical high-precision calculations also 
for non-relativistic bound states in NRQED/NRQCD.

\section{pNRQED/pNRQCD}
\label{secpNR}
\index{pNRQCD} 
\index{pNRQED}
The problem of disentangling the soft from the ultrasoft scale 
in NRQED/NRQCD was addressed immediately after dimensional regularization was established as 
an useful tool for non-relativistic bound state calculations also.
The history and details of the developments that have ultimately led to the construction 
of EFTs for the ultrasoft degrees of freedom of NRQED/NRQCD have been recollected 
in Ref. \refcite{Brambilla:2004jw} and we refer the interested reader to it. 
Here, we would like just to stress the importance that the process of $t\bar{t}$ production near threshold 
(see  Ref. \refcite{Brambilla:2004wf} and references therein)
has played in these developments, providing the only  near threshold, heavy quark-antiquark 
system in nature entirely accessible in perturbation theory. As it was mentioned before, this very special 
feature of the $t\bar{t}$ system near threshold had already been appreciated by the groups working 
on the subject at the beginning of the nineties and, in particular, by the Vienna group.

In the following, in order to illustrate some general features, 
we will concentrate on the EFTs for ultrasoft degrees of freedom of NRQED/NRQCD 
known as potential NRQED \cite{Pineda:1998ie,Pineda:1998kn} and potential NRQCD \cite{Pineda:1998bj,Brambilla:1999xf}
(for an alternative formulation see Ref. \refcite{Luke:1999kz} and the review in Ref. \refcite{Hoang:2002ae}).
The large scale that is integrated out in pNRQED/pNRQCD is the typical momentum transfer of the bound 
state, which, in coordinate space, is associated with the inverse of the typical distance $r$ 
between the two heavy particles. The degrees of freedom of  pNRQED/pNRQCD are non-relativistic fermions 
and antifermions, and photons/gluons of energy and momentum smaller than $mv$. 
They build up the operators $O_{k,n}$ of the Lagrangian; the operators may be also chosen to be explicitly gauge invariant. 
The Lagrangian is organized as an expansion in $1/m$, inherited from NRQED/NRQCD, and an 
expansion in $r$ (multipole expansion), which is characteristic of the new EFT:
\be
{\cal L}_{\rm pNRQED/pNRQCD} = 
\sum_{k,n} \frac{1}{m^k} \times c_k(\als(m),\mu) \times V_n(r,\mu,\mu^\prime)  
\times r^n \, O_{k,n}(\mu^\prime).
\label{pNRQCD}
\ee
Since, once $O_{k,n}$ has been run down to the lowest energy $mv^2$, the expectation 
value of $O_{k,n}$ scales like $mv^2$, Eq. (\ref{pNRQCD}) provides, for any physical observable,  
a perturbative expansion in the ratio of $mv^2$ over $mv$ or $m$.
The Wilson coefficients $c_k$ are those inherited from NRQED/NRQCD, the Wilson coefficients $V_n$ 
are the new ones of pNRQED/pNRQCD. They are non analytical in the scale $r$
and function of the new factorization scale $\mu^\prime$.
They are calculated by matching, order by order in $r$, Green's function in NRQED/NRQCD with Green's function 
in pNRQED/pNRQCD. In pNRQED, the matching may be also done      
order by order in $\alpha$. In pNRQCD, $\als(mv) \ll 1$ holds only for tightly bound states 
(short-range quarkonia, e.g. the bottomonium ground state or $t\bar{t}$ near threshold), 
while, in general, higher excited quarkonium states (long-range quarkonia)
are not accessible by perturbation theory. This means that we can rely on an expansion in  $\als(mv)$ only 
for the former states, while for the latter states the matching has to be done in a non-perturbative fashion.

\begin{figure}[htb]
\makebox[0cm]{\phantom b}
\put(0,0){{\epsfxsize=11truecm\epsffile{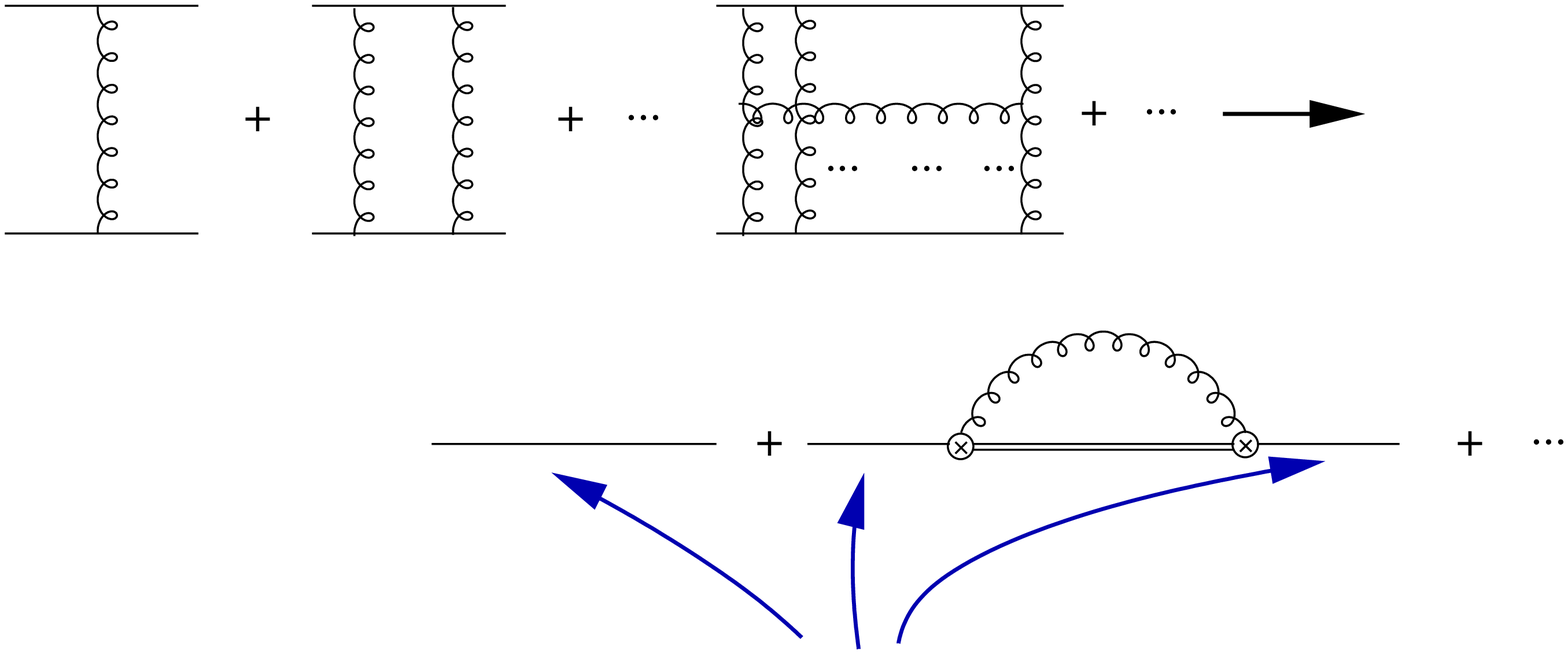}}}
\put(75,65){NRQCD}
\put(75,10){pNRQCD}
\put(175,-15){$\displaystyle \frac{1}{E - p^2/m- V(r,\mu,\mu^\prime)}$}
\caption{Matching to pNRQCD.
\label{figpNRQCD}}
\end{figure}

Let us consider the NRQCD diagram shown in Fig. \ref{figNRQCD} and the part of it where soft gluons are 
exchanged between the quark and antiquark.  The sum of all soft-gluon exchanges would be described in 
pNRQCD by the diagram shown on the right in Fig. \ref{figpNRQCD} where the single line 
stands for a quark-antiquark propagator in a color-singlet configuration, ${1}/{(E - p^2/m- V)}$, the double line for 
a quark-antiquark pair in a color-octet configuration, the curly line for ultrasoft gluons and 
the circle with a cross for a chromoelectric dipole interaction $\sim \phi^\dagger {\bf r}\cdot {\bf E}\,\phi$ 
that comes from multipole expanding the gluon fields in the NRQCD Lagrangian.
The non-analytical behaviour in $r$ of the NRQCD diagram is reproduced in pNRQCD by the 
Wilson coefficient $V$. Since $V$, together with $p^2/m$, makes up the pole of the quark-antiquark 
propagator, the interpretation of $V$ is obvious: $V$ is the potential describing the interaction 
in the heavy quark-antiquark pair. 
At leading order in the multipole expansion, when we neglect diagrams involving 
ultrasoft gluons, the equation of motion of a non-relativistic 
fermion-antifermion pair is nothing else than the Schr\"odinger equation (\ref{schroephi}). 

\index{Schr\"odinger equation}
The Schr\"odinger equation is the equation governing non-relativistic bound states 
in quantum mechanics. The full relativistic description provided by field theory, which is richer and much more 
complex, is given by the Bethe--Salpeter equation. This complexity arises 
from the entanglement of different energy scales. Once the contributions of all these 
scales have been separated/factorized, we are left with an EFT of the ultrasoft degrees of freedom.
The Schr\"odinger equation naturally emerges as the equation of motion of these  
ultrasoft degrees of freedom.
But, because the EFT contains all the richness and complexity of the field theory, although 
unfolded in a systematic and organized way, 
the Schr\"odinger equation, which we have gotten from the EFT, 
is much more than the Schr\"odinger equation of quantum mechanics 
we have started with. First, the EFT provides a proper, field theoretically founded, 
definition of the potential: the potential is the Wilson coefficient of the dimension six operator 
of the EFT, containing two fermion and two antifermion fields,  
that encodes all contributions coming from modes whose energies and momenta are larger than the binding energy. 
It undergoes renormalization, develops scale dependence and satisfies renormalization
group equations, which, in perturbation theory, allow to resum potentially large logarithms.
Moreover, the EFT accounts also for effects that cannot be cast in a Schr\"odinger equation 
and that are due to the coupling of the fermion-antifermion pair with the other ultrasoft degrees of freedom. 

\begin{figure}[htb]
\makebox[0cm]{\phantom b}
\put(100,-50){\epsfxsize=5truecm \epsffile{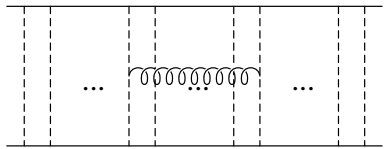}}
\caption{QCD diagrams responsible for the infrared sensitivity of the static potential.
\label{figADM}}
\end{figure}

\index{potential in QCD}
In QCD, ultrasoft effects affect also the static potential. As first observed 
in Ref. \refcite{Appelquist:1978es}, they come from the 
``non-Abelian Lamb shift''\cite{Kummer:1996jz} diagrams displayed in Fig. \ref{figADM}.
At fixed order in perturbation theory, the diagrams are infrared divergent; at order $\als^4$, the 
leading logarithmic correction is \cite{Brambilla:1999qa}
\be
\delta V(r,\mu^\prime)=
-\frac{3}{r}\frac{\als(\mu^\prime)}{\pi} \als^3(1/r) \ln ({r\mu^\prime}).
\label{Vlog}
\ee
The result shows clearly the non-physical nature of the potential, which 
depends on the renormalization scale $\mu^\prime$. 
The potentially large logarithms, $\ln ({r\mu^\prime})$, have been resummed by means of renormalization 
group equations in Ref. \refcite{Pineda:2000gz}; subleading corrections have been calculated in Ref. 
\refcite{Brambilla:2006wp}. In physical observables, like the static energy or the quarkonium mass, 
the scale dependence of Eq. (\ref{Vlog}) cancels against ultrasoft contributions coming from the second diagram 
in the pNRQCD part of Fig. \ref{figpNRQCD}.

\index{relativistic invariance}
Higher-order terms in the relativistic expansion may be computed systematically in the EFT.
Again, the full complexity and symmetries of the underlying field theory are not lost in the expansion.
So, for instance, relativistic invariance imposes specific constraints on the Wilson coefficients/potentials 
of the EFT \cite{Brambilla:2001xk,Brambilla:2003nt}, which can be tested on the lattice \cite{Koma:2006fw,Koma:2007jq}.

\index{quarkonium}
\index{positronium}
Applications of pNRQCD and, more in general, of EFTs for the ultrasoft degrees
of freedom of NRQCD have led to a plethora of new results in quarkonium physics  
(see Refs. \refcite{Brambilla:2004jw,Brambilla:2004wf,Vairo:2006nq,Soto:2006zs,Vairo:2006pc,Brambilla:2007vr}
for some recent reviews) and, in particular, in $t\bar{t}$ threshold production 
(see Refs. \refcite{Hoang:2001mm,Hoang:2003xg,Hoang:2006pc,Pineda:2006ri} 
for the present status of the art).
Also QED calculations have remarkably benefitted from the EFT approach and corrections of very 
high order in perturbation theory have been calculated in the last years for many observables 
after decades of very slow or no progress. As an example, we mention that for the 
hyperfine splitting of the positronium ground state the terms of order $\alpha^6$, $\alpha^7 \ln^2\alpha$ and 
$\alpha^7\ln \alpha$ have been calculated (for recent reviews on positronium precision studies and further references 
we refer to Refs. \refcite{Karshenboim:2003vs,Penin:2003jz}). 

\begin{figure}[htb]
\makebox[2cm]{\phantom b}
\put(0,0){\epsfxsize=7truecm\epsffile{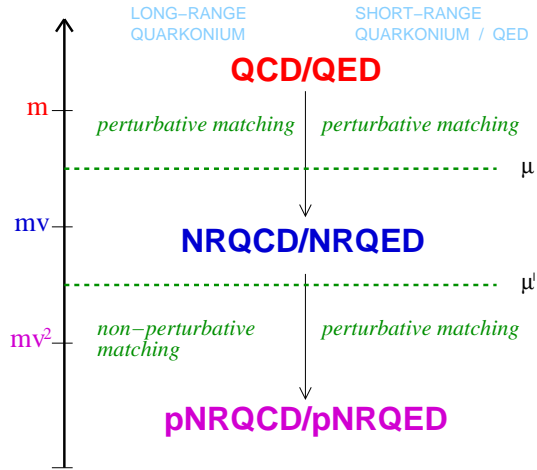}}
\caption{EFTs for bound states in QED and, for heavy quarks, in QCD.\vspace{-0.1mm}}
\label{fig:scales}       
\end{figure}

In Fig. \ref{fig:scales}, we summarize the hierarchy of EFTs for bound states in QED 
and, for heavy quarks, in QCD.

\section{Outlook}
The history of non-relativistic bound states in the quantum theory 
had in the last century a peculiar spiral behaviour. It started with the 
Schr\"odinger equation of the hydrogen atom and seemed to have written 
its ultimate chapter with the Bethe--Salpeter equation in the fifties. 
However, in face of the enormous difficulties in treating bound states in field theory 
by means of the Bethe--Salpeter equation, a long journey started in the seventies 
that took us back to the Schr\"odinger equation.
This coming back, however, was not like closing a circle, it was more like building up a spiral.
The Schr\"odinger equation we have come back to, encompasses 
all the complexity of the Bethe--Salpeter equation, all the richness of field 
theory, in the elegant and systematic setting of non-relativistic effective 
field theories. The counting rules and structure of the EFTs have allowed 
us to perform calculation with unprecedented precision, where 
higher-order perturbative calculations were possible, and to systematically 
factorize short from long range contributions where observables 
were sensitive to the non-perturbative, infrared dynamics of QCD.

Non-relativistic EFTs have become nowadays the standard tool to treat 
non-relativistic bound states. Besides QED bound states and quarkonium, these include 
hadronic atoms like pionium \cite{Gasser:2007zt}, 
nucleon-nucleon systems \cite{Beane:2000fx,Epelbaum:2008ga}, non-relativistic 
bound states at finite temperature
\cite{Brambilla:2008cx,Escobedo:2008sy,Laine:2008cf} and many others.
The modern history of non-relativistic bound states is far from being fi\-nished
and still needs to be told in its full extent.
\vspace{-3mm}

\section*{Acknowledgements}
I thank Andr\'e Hoang for comments.
I acknowledge financial support from the European Research Training Network 
FLAVIA{\it net} (FP6, Marie Curie Programs, Contract MRTN-CT-2006-035482) and 
from the DFG cluster of excellence "Origin and structure of the universe" (www.universe-cluster.de).
\vspace{-2.5mm}

\bibliographystyle{ws-rv-van}
\bibliography{KummerMemorial_hep}

%\printindex[aindx]                 % to print author index
%\printindex                         % to print subject index
\end{document}